 \documentstyle[aps,preprint,prl]{revtex}

\begin{document}
\bibliographystyle{prsty}
\draft

\title{Born effective charges of Barium Titanate: band by band decomposition
and
 sensitivity to structural features}
\author{Ph. Ghosez$^{\ast}$, X. Gonze$^{\ast}$, Ph. Lambin$^{\dagger}$
 and J.-P. Michenaud$^{\ast}$}
\address{$^{\ast}$Unit\'e PCPM,
Universit\'e Catholique de Louvain, B-1348 Louvain-La-Neuve, Belgium \\
$^{\dagger}$D\'epartement de Physique, Facult\'es Universitaires Notre
Dame de la Paix, B-5000 Namur, Belgium}
\date{\today}
\maketitle
\begin{abstract}
The  Born effective charge tensors of Barium Titanate have
been calculated for each of its 4 phases.
Large effective charges of Ti and O, also predicted by
shell model calculations and made plausible by a simplified model,
reflect the partial covalent
character of the chemical bond.  A band by band decomposition
confirms that orbital hybridization is not restricted to Ti and
O atoms but also involves Ba which appears more covalent than
generally assumed. Our calculations reveal a strong dependence of
the effective charges on the atomic positions contrasting with a
relative insensitivity on isotropic volume changes.

\end{abstract}

 \newpage


Barium Titanate (BaTiO$_{3}$) is a well known ferroelectric compound
\cite{Lines}.
Its structure, cubic perovskite at
high temperature, becomes tetragonal around 120$^\circ$C, orthorhombic at
about 0$^\circ$C
and rhombohedral near -70$^\circ$C. Although it is probably one of
the most studied ferroelectric crystal \cite{Lines},
the mechanism of its phase transitions is
still subject of controversy \cite{Mario}.
The character of the chemical bonding also remains questionable \cite{Cohen}.

The Born effective charge $Z^{*}$ is a
fundamental quantity for the study of lattice dynamics, controlling
the long range Coulomb part of the force constants.
An estimation of $Z^{*}$ for three ABO$_{3}$ compounds
was first proposed by Axe \cite{Axe}
on the basis of a phenomenological theory. However,
the crude hypothesis inherent in his procedure only allowed an approximate
estimation
of these charges and restricted his investigations to the cubic phase. New
advances in ``ab initio'' techniques now enable one to determine $Z^{*}$
theoretically
using perturbation theory \cite{GonzeRF} or finite difference of
polarisation \cite{KSV2}.
Accurate values have been recently reported by Resta {\it et al.} for
KNbO$_{3}$ \cite{RestaPRL},
by ourselves for BaTiO$_{3}$ \cite{GGM} and
by Zhong {\it et al.} \cite{Zhong1} for a whole set of ABO$_3$ compounds.
These studies underline the surprisingly large values already observed by Axe
and  generally referred to as ``anomalous'' charges.
For BaTiO$_3$, no attempt was performed to determine how these giant charges
are affected by structural details, even if this information could reveal
essential
for an accurate investigation of lattice dynamics. We will show that $Z^{*}$
are
relatively insensitive to isotropic volume changes but are strongly
affected by changes
of positions associated to the phase transitions.

Moreover, if obtaining the values of $Z^{*}$ is already an important result,
to understand why some charges are so large also constitutes a fondamental
problem.
Until now, anomalous charges in ABO$_3$ materials were explained qualitatively
in the framework of a bond orbital model, following Harrison \cite{Harrison}.
Recently, Posternak {\it et al.} \cite{Posternak} elegantly emphazised the
role of
covalency between Nb and O in KNbO$_{3}$ from the analysis of giant
effective charges.
We show here that, in a more general way, a band by band
decomposition of $Z^{*}$ is a sensitive tool to investigate the role
of covalency and ionicity without any preliminary hypothesis on the
orbitals that interact.
Our study helps to clarify the debate on the real nature of the chemical
bonding
in BaTiO$_{3}$ and brings out the role of the Ba atom.


We work in the framework of the Density
Functional Formalism within the local density
approximation \cite{JG}. The Born effective charge tensor
$Z_{\kappa,\gamma \alpha}^{*}$ of atom $\kappa$ can be linked either to
the change of polarisation $P_{\gamma}$ induced by the
periodic displacement $\tau_{\kappa,\alpha}$,
or to the force $F_{\kappa,\alpha}$ induced on atom $\kappa$ by
an electric field $\cal{E}_{\gamma}$:
$
Z_{\kappa,\gamma \alpha}^{*}
=  V \frac{\partial P_{\gamma}}{\partial \tau_{\kappa,\alpha}}
=  \frac{\partial F_{\kappa,\alpha}}{\partial \cal{E}_{\gamma}}
= - \frac{\partial^{2} E}{\partial \cal{E}_{\gamma}
 \partial \tau_{\kappa,\alpha}}
$.
So, it also  appears
as a mixed second derivative of the total energy $E$ per unit cell volume V,
which can be evaluated using the variational approach to
Density Functional Perturbation Theory \cite{GonzeRF,GGM}.
We use the exchange-correlation energy functional and the
norm-conserving pseudopotentials of  ref. \cite{GGM}.
The Brillouin Zone was sampled with a 6$\times$6$\times$6
{\it k}-point set.
The plane-wave basis set was determined by a 35 hartrees energy cutoff
which guarantees a convergence  better than 0.5\% on $Z^{*}$.



First we investigate the Born effective charges in the cubic structure.
Ab-initio values, obtained at the experimental (a$_{cell}$=4\AA) and
theoretically optimized (a$_{cell}$=3.94\AA) lattice parameter,
as well as those corresponding to a compressed cubic cell (a$_{cell}$=3.67\AA)
are reported in Table ~\ref{Zcubic}.
In comparison, we also computed the effective charges from the
shell model parameters proposed by Khatib {\it et al.} \cite{Khatib}.

The charges of Ba and Ti, isotropic owing to symmetry,
 are equal respectively (at the theoretical volume) to  +2.77 and +7.25.
For O, the values of $Z^{*}_{O_\parallel}$ (-5.71) and
$Z^{*}_{O_\perp}$ (-2.15) refer respectively to
a displacement of the oxygen ion along the Ti-O direction or
perpendicular to it.
Our results are in good agreement with those obtained by Zhong {\it et al.}
\cite{Zhong1} (also at the theoretical volume) from
finite difference of polarisation and
globally reproduce the values deduced by Axe \cite{Axe}.
The accordance with the shell model is only qualitative and
illustrates the limited precision obtained within such an empirical
description.
We note however that this shell model was not designed to reproduce the
dielectric
properties of BaTiO$_{3}$. The large value of $Z^{*}_{Ti}$ (7.51) proves that
it
implicitely includes covalency effects responsible for
anomalous effective charges, as described below.

The charges on Ti and O$_\parallel$ are surprisingly large in
the sense that they reach about twice the value they would
have in a pure ionic picture: they reveal the presence
of a large dynamic contribution superimposed to the static charge.
As the latter quantity is ill defined, determining the dynamic
contribution is an ambiguous task. So, in the following,
we prefer to speak in terms of {\it anomalous} contributions that we define
as the additional charge with respect to the well known, ionic
value (cfr. first column of Table ~\ref{Zcubic}).

The physical possibility of obtaining anomalous charges  can be understood
within a very simplified model. Let us just consider
a diatomic molecule XY with an interatomic distance $u$ and
a dipole moment $P(u)$. The dipole moment enables us to {\it define} a static
charge $Z(u)=\frac{P(u)}{u}$ and a dynamic charge
$Z^{*}(u)= \frac{\partial P(u)}{\partial u}$ also equal to
$Z(u)+u\frac{\partial Z(u)}{\partial u}$.
As the distance between X and Y is modified from 0 to some $\overline{u}$
(the distance corresponding to a complete transfer of electrons from X to Y),
the dipole moment evolves continuously from $P(0)=0$ (since there is no
dipole for that case)
to $P(\overline{u})$.
Interestingly,
$ \int^{\overline{u}}_{0} Z^{*}(u) du =
[P(\overline{u})-P(0)] = \overline{u} Z(\overline{u})$.
So, $\frac{1}{\overline{u}}\int^{\overline{u}}_{0} Z^{*}(u) du =
Z(\overline{u})$:
the mean value of $Z^{*}(u)$ from 0 to $\overline{u}$ is equal to
$Z(\overline{u})$.
This result guarantees that, if $Z(u)$ changes with $u$,
$Z^{*}(u)$ has to be greater than $Z(\overline{u})$
for some $u$ between $[0, \overline{u}]$. The
difference between $Z(u)$ and $Z^{*}(u)$ can be large if $Z(u)$ changes
rapidly with $u$.

In BaTiO$_{3}$, the approximate reciprocity between O$_{\perp}$ (-3.71) and
Ti (+3.25)
anomalous contributions suggests that they correspond to a global transfer of
charge from
O to Ti when the Ti-O distance shortens. In the framework
of the bond orbital model proposed by Harrison \cite{Harrison}, the charge
redistribution
is attributed to a change in the hopping integral produced by dynamic
modification of orbital hybridizations. Matching this bond orbital model to
the real material
ask for the identification of the relevant orbitals.

Hybridization between O 2{\it p} and the unoccupied metal {\it d} orbitals
is well known
and was already pointed out from experiments \cite{exp},
tight-binding models \cite{Harrison} and LCAO calculations \cite{Michel}.
It was highligthed more recently by Cohen \cite{Cohen} from
first-principles as an essential feature of ABO$_3$ ferroelectric compounds.
In this context, it seemed realistic, following Harrison, to focus on
O 2{\it p} -Ti 3{\it d} hybridization changes to explain intuitively large
anomalous contributions \cite{Zhong1}. Posternak {\it et al.}
\cite{Posternak} went beyond this
credible assumption. They showed for KNbO$_3$ that the anomalous contributions
disappear when the interaction between the O 2{\it p} and Nb 4{\it d}
orbitals is
artificially suppressed.

Nothing proves that the conclusions would be the same for BaTiO$_3$ neither
that the hybridizations are limited to this kind of {\it p-d} interactions.
Theoretical investigations, confirmed by experimental results \cite{exp},
suggest
that Ba also plays a major role in forming the valence band structure.
Now, we propose
a more detailed investigation, based on a band by band analysis of
$Z^{*}$\cite{GGM2}.


Contrariwise to the total effective charge tensor, the band by band
decomposition
depends not only on the Hilbert space of occupied valence states but
also on the particular valence eigenfunctions.
In order to identify the contribution of each band we  worked in the
``diagonal gauge'' \cite{XG} in which the matrix of the first-order
eigenenergies
with respect to atomic displacements is diagonal.
This is equivalent to associate Wannier functions with each
separate set of bands for calculating the polarization \cite{KSV2,VKS}.

For a {\it reference} configuration in which the contribution to the charge
of a given atom
is -2 for each band associated to its own orbitals and 0 for the other
bands, each Wannier
function is centered on a given atom.
Within the recent theory of the polarization \cite{KSV2,VKS},
the {\it anomalous charge} associated to a particular band for a given ion
(defined as the additional part with respect to our {\it reference} value)
reflects how the center of the Wannier function of this band moves with
respect to the ion.
In a purely ionic crystal, each band would be composed of non hybridized
orbitals. As covalency develops, different orbitals mix
adding anomalous contributions from several bands.

Results for the theoretical cubic structure are reported in
Table~\ref{Zbbb}. The
first line brings together the charge of the nucleus and the core electrons
included in the pseudopotential. The other contributions come from the valence
electron levels. Sets of bands were identified by the name of the
main atomic orbital which generated this energy level in the solid
(Fig.~\ref{Fbbb}).
Their dominant character was confirmed by partial density plots.
As a check of the coherency of our decomposition, the sum of the contributions
of each fully occupied band, obtained independently, is -2
(within $\pm 0.01$, the accuracy of our calculation).

As expected, the main $Z^{*}_{Ti}$
anomalous charge is localized on the O 2{\it p}
bands (+2.86). It can be understood
by an hybridization between O 2{\it p} and Ti 3{\it d} orbitals.
Interestingly, there are also smaller but non-negligible anomalous
charges from the Ti 3{\it p} (-0.22),  O 2{\it s} (+0.23) and
Ba 5{\it p} (+0.36) bands.
The different positive contributions correspond to a
displacement of the center of the Wannier
function of the O and Ba bands in the direction of Ti when this atom moves.
The Ti3{\it s} contribution is close to -2.
This result {\it a posteriori} justifies the inclusion of deeper
levels as part of ionic pseudopotentials.

For Z$^{*}_{Ba}$, the decomposition
is more surprising: the small global anomalous effective charge (+0.77)
that could be typical of a more ionic character appears partially fortuitous.
The anomalous charges of O 2{\it s} (+0.73) and O 2{\it p} (+1.50) bands
are {\it not} small.
Nevertheless, they are partially compensated by Ba 5{\it p} (-1.38) and
Ba 5{\it s} (-0.11) anomalous contributions.
This result gives a tangible proof of the hybridization of Ba 5{\it p} orbitals
already suggested from the experiment \cite{exp}.

Concerning oxygen, even if O$_\parallel$ and O$_\perp$ are defined
respectively for
a displacement of O
in the Ti and Ba direction, it seems only qualitative to associate
$Z^{*}_{O_\parallel}$
with $Z^{*}_{Ti}$ and $Z^{*}_{O_\perp}$ with $Z^{*}_{Ba}$.
The O 2{\it p} anomalous contributions to Ti and O$_\parallel$ do not
compensate.
Moreover, O 2{\it p} contribution to $Z^{*}_{Ba}$ not only comes from O$_\perp$
but has equivalent contributions coming from O$_\parallel$.

Within this analysis, several bands appear as a complex mixing of orbitals
coming from the different ions. In this context, and contrasting with the
conclusion
of Posternack {\it et al.} for KNbO$_{3}$, a correct understanding of the
Born effective charge goes here beyond the simple model of Harrison.
Our result clarifies the mixed ionic-covalent character of BaTiO$_3$:
it clearly establishes that the covalent character is not restricted
to the Ti-O bond as generally assumed.


Until now, calculations of $Z^{*}$ essentially focused on the cubic phase
\cite{GGM,Zhong1}.
On the basis of an investigation of these charges in the experimental
tetragonal structure
of KNbO$_3$ \cite{RestaPRL} and PbTiO$_3$ \cite{Zhong1}, it was argued that
they are quite insensitive
to structural details.
This result is surprising if we remember that anomalous contributions to
$Z^{*}$
are closely connected to orbital hybridizations, these in turn, being strongly
affected by phase transitions \cite{Cohen}. In addition, the theoretical
overestimation of the
spontaneous polarization for the rhombohedral structure of BaTiO$_3$
\cite{Zhong1} also
suggests a reduction of $Z^{*}$ in this phase.

We computed the Born effective charge tensors for the 3 ferroelectric phases at
the experimental unit cell parameters \cite{cell},
with relaxed atomic positions \cite{GGLM}.
In this paper, we only comment on the eigenvalues of these tensors
(Tables~\ref{ZBaTi}-~\ref{ZOx})
that already allow a pertinent comparison with the cubic phase.
The $Z^{*}_{33}$ eigenvalues of Ba and Ti correspond to an eigenvector aligned
along the ferroelectric axis. In the case of O, the eigenvector
associated to the highest eigenvalue approximately point in the Ti-O
direction: we identify
this highest contribution as O$_{\parallel}$ while the others are refer as
O$_{\perp}$,
by analogy with the cubic phase.

Although the charges of Ba and O$_{\perp}$ remain globally the same in the
4 phases,
for Ti and O$_{\parallel}$, stronger modifications are observed.
Changing the Ti position by 0.076\AA\ (2\% of the unit cell length) when
going from
the cubic to the rhombohedral phase, reduces the {\it anomalous}
part of $Z^{*}_{Ti}$ {\it by more than 50\%} along the ferroelectric axis
(Table~\ref{ZBaTi}).
The amplitude of $Z^{*}_{Ti}$ and $Z^{*}_{O}$ in the direction of the shortest
Ti-O bond length $d_{min}$ of each phase is plotted in Fig.~\ref{FZTiO}
with respect to the interatomic distance $d_{min}$.
For the different phases at zero pressure, the anomalous parts decrease
with $d_{min}$.
The comparison with a compressed cubic phase at 3.67 \AA\ in which the Ti-O
distance is {\it the same} that the shortest Ti-O bond length
in the tetragonal structure shows nevertheless that the evolution of
$Z^{*}_{Ti}$  cannot
be explained in terms of the Ti-O distance only but is critically affected
by the anisotropy
of the Ti environnement.

This is clear from the band by band decompositions of $Z^{*}_{Ti}$ in
Table~\ref{Zbbb}.
While in the cubic structure at 3.67 \AA\
every Ti-O distance is equivalent to the others, in the tetragonal phase, along
the ferroelectric axis, a short Ti-O bond length is followed by a larger
one which breaks the
Ti-O chain in this direction and inhibits the giant current associated to
the large effective charges.
This appears at the level of the O 2{\it p} bands (+1.48 instead of +2.86)
while the other contributions remains
equivalent to those of the cubic phase at 3.94 \AA. Analysis from the cubic
structure
at 3.67 \AA\ reveals that the O 2{\it p} contribution is not significantly
affected by hydrostatic
pressure; on the other hand,  the anomalous parts of the Ba 5{\it p}, Ba
5{\it s } and Ti 3{\it p}
bands are modified by about 50 \% due to the compression.


In this work, we were able to compute the Born effective charges
of BaTiO$_{3}$ in its 4 phases.
Effective charges are a sensitive tool for analysing  dynamic changes
of orbitals hybridizations, especially if a band by band decomposition is
performed.
In our description Ba appears more covalent than generally assumed. The charges
of Ti and O are strongly affected by atomic
displacements but quite insensitive to hydrostatic pressure.

We thank Z. Levine, D. C. Allan for sharing some of their routines
and J.-M. Beuken for informatic support.
X.G. and Ph.L. acknowledge financial support from the FNRS-Belgium.
We are grateful to Corning Inc. (M.P. Teter and D.C. Allan)
for the availability of the PlaneWave code
for ground-state calculations. We used IBM-RS 6000 work
stations from common projects between IBM-Belgium,
UCL-PCPM and FUNDP.


\begin{figure}

\caption{Electronic band structure of cubic BaTiO$_3$ (a$_{cell}$=3.94\AA).}
\label{Fbbb}
\end{figure}

\begin{figure}
\caption{Born effective charge of Ti (opened symbols) and O (filled symbols)
in the direction of the shortest Ti-O bond length $d_{min}$ with respect to
this interatomic distance
for the cubic (square), tetragonal (lozenge),
orthorhombic (circle) and rhombohedral (triangle)
phases.}
\label{FZTiO}
\end{figure}

\begin{table}
\caption {Born effective charges for
cubic BaTiO$_3$. }
\label{Zcubic}
\begin {tabular}{lccccccc}
&Ions &Axe\cite{Axe} &Shell Model  &Zhong {\it et al.}\cite{Zhong1}
&a$_{cell}$=4.00\AA &a$_{cell}$=3.94\AA  &a$_{cell}$=3.67\AA   \\
\hline
$Z^{*}_{Ba}$            &+2       &2.9    &1.63     &2.75     &2.74
&2.77    &2.95  \\
$Z^{*}_{Ti} $           &+4       &6.7    &7.51     &7.16     &7.29
&7.25    &7.23  \\
$Z^{*}_{O_{\perp}}$     &-2       &-2.4   &-2.71    &-2.11    &-2.13
&-2.15   &-2.28 \\
$Z^{*}_{O_{\parallel}}$ &-2       &-4.8   &-3.72    &-5.69    &-5.75
&-5.71   &-5.61 \\
\end{tabular}
\end{table}

\narrowtext

\begin{table}
\caption {Band by band decomposition of $Z^{*}$ (see text).}
\label{Zbbb}
\begin {tabular}{lcccccc}
&\multicolumn{4}{c}{cubic(3.94\AA)}
&cubic(3.67\AA)  &tetragonal \\
Band &$Z_{Ba}^{*}$  &$Z_{O_{\perp}}^{*}$
 &$Z_{O_{\parallel}}^{*}$
 &$Z_{Ti}^{*}$    &$Z_{Ti}^{*}$ &$Z_{Ti}^{*}$ \\
\hline
core    &10.00    &6.00   &6.00  &12.00   &12.00  &12.00 \\
Ti 3{\it s}  &0.01   &0.00  &0.02  &-2.03        &-2.07  &-2.05 \\
Ti 3{\it p}  &0.02    &-0.02  &0.21  &-6.22      &-6.43  &-6.26 \\
Ba 5{\it s}  &-2.11    &0.02  &0.01  &0.05       &0.09   &0.05  \\
O 2{\it s}   &0.73    &-2.23  &-2.51  &0.23     &0.27   &0.25  \\
Ba 5{\it p}  &-7.38    &0.58  &-0.13  &0.36      &0.64   &0.34  \\
O 2{\it p}   &1.50    &-6.50  &-9.31   &2.86    &2.73   &1.48  \\
\hline
Total        &2.77    &-2.15  &-5.71  &7.25  &7.23  &5.81\\
\end{tabular}
\end{table}

\begin{table}
\caption {Eigenvalues of the Born effective charge tensors of Ba and Ti
at the experimental volumes.}
\label{ZBaTi}
\begin {tabular}{lcccccccc}
phase  &Z$^{*}_{Ti,11}$  &Z$^{*}_{Ti,22}$ &Z$^{*}_{Ti,33}$  & &
&Z$^{*}_{Ba,11}$  &Z$^{*}_{Ba,22}$ &Z$^{*}_{Ba,33}$ \\
\hline
cubic           &7.29  &7.29  &7.29  & & &2.74  &2.74  &2.74    \\
tetragonal      &6.94  &6.94  &5.81  & & &2.72  &2.72  &2.82    \\
orthorhombic    &6.80  &6.43  &5.59  & & &2.72  &2.81  &2.77    \\
rhombohedral    &6.52  &6.52  &5.58  & & &2.79  &2.79  &2.74    \\
\end{tabular}
\end{table}

\widetext

\begin{table}
\caption {Eigenvalues of the Born effective charge tensors of O
at the experimental volumes.}
\label{ZOx}
\begin {tabular}{lcccccccccc}
&\multicolumn{6}{c}{O$_{\perp}$} & &\multicolumn{3}{c}{O$_{\parallel}$} \\
phase  &Z$^{*}_{O1,11}$  &Z$^{*}_{O2,11}$ &Z$^{*}_{O3,11}$
&Z$^{*}_{O1,22}$  &Z$^{*}_{O2,22}$ &Z$^{*}_{O3,22}$
& &Z$^{*}_{O1,33}$  &Z$^{*}_{O2,33}$ &Z$^{*}_{O3,33}$ \\
\hline
cubic           &-2.13  &-2.13  &-2.13  &-2.13  &-2.13  &-2.13 & &-5.75
&-5.75  &-5.75  \\
tetragonal      &-1.99  &-1.95  &-1.95  &-1.99  &-2.14  &-2.14 & &-4.73
&-5.53  &-5.53  \\
orthorhombic    &-1.91  &-1.91  &-1.97  &-2.04  &-2.04  &-2.01 & &-4.89
&-4.89  &-5.44  \\
rhombohedral    &-1.97  &-1.97  &-1.97  &-1.97  &-1.97  &-1.97 & &-5.03
&-5.03  &-5.03  \\
\end{tabular}
\end{table}


\vspace{2mm}






\begin{thebibliography}{10}
\bibitem{Lines}
M. E. Lines and A.M. Glass, ``Principles and Applications of Ferroelectrics
and Related Materials.'', Clarendon Press, Oxford (1977).
\bibitem{Mario}
K. Laabidi et al., Europhys. Lett. 26, 309 (1994).
\bibitem{Cohen}
R. E. Cohen, Nature 358, 136 (1992).
\bibitem {Axe}
J. D. Axe, Phys. Rev. 157, 429 (1967).
\bibitem{GonzeRF}
X. Gonze, D.C. Allan and M.P. Teter, Phys. Rev. Lett. 68, 3603 (1992).
\bibitem{KSV2}
R.D. King-Smith and D. Vanderbilt, Phys. Rev. B 47, 1651 (1993).
\bibitem{RestaPRL}
R. Resta, M. Posternak and A. Baldereschi, Phys. Rev. Lett. 70, 1010 (1993).
\bibitem{GGM}
Ph. Ghosez, X. Gonze and J.-P. Michenaud, Ferroelectrics 153, 91 (1994).
\bibitem{Zhong1}
Zhong, R.D. King-Smith and D. Vanderbilt, Phys. Rev. Lett. 72, 3618 (1994).
\bibitem{Harrison}
W. A. Harrison, ``Electronic Structure and the Properties of Solids,
Dover, New-York (1980).
\bibitem {Posternak}
M. Posternak, R. Resta and A. Baldereschi, Phys. Rev. B 50, 8911 (1994).
\bibitem{JG}
R. O. Jones and O. Gunnarsson, Rev. Mod. Phys. 61, 689 (1989).
\bibitem{Khatib}
D. Khatib, R. Migoni, G. E. Kugel and L. Godefroy, J. Phys.: Condens. Matt.
1, 9811 (1989).
\bibitem{exp}
V. V. Nemoshkalenko and A. N. Timoshevskii, Phys. Stat. Sol. b127, 163 (1985);
L. T. Hudson, R. L. Kurtz, S. W. Robey, D. Temple and R. L. Stockbauer,
Phys. Rev. B 47, 1174 (1994).
\bibitem{Michel}
F. M. Michel-Calendini, H. Chermette and J. Weber, J. Phys. C: Sol. St. Phys.
13, 1427 (1980).
\bibitem{GGM2}
Ph. Ghosez, X. Gonze and J.-P. Michenaud, proceedings of the ``Third
Williamsburg
Workshop on First Principles Calculations for Ferroelectricity'' (1994).
\bibitem{XG}
X. Gonze, unpublished.
\bibitem{VKS}
D. Vanderbilt and R. D. Kingh-Smith, Phys. Rev. B 48, 4442 (1993).
\bibitem{cell}
A. W. Hewat, Ferroelectrics 6, 215 (1974); G. H. Kwei, A. C. Lawson, S. J.
L. Billinge
and S.-W. Cheong, J. Phys. Chem. 97, 2368 (1993).
\bibitem{GGLM}
Ph. Ghosez, X. Gonze, Ph. Lambin and J. P. Michenaud, unpublished.
\end{thebibliography}
\end{document}